\documentclass{aa}

\begin{document}

%
%
\thesaurus{07(07.16.8; 07.16.09; 07.16.06; 13.25.3)}

\title{A Search for X-ray emission from Saturn, Uranus and Neptune}

\author{Jan-Uwe Ness \and J\"urgen H.M.M. Schmitt} 
\authorrunning{J.U. Ness \& J.H.M.M. Schmitt}
\titlerunning{X-ray emission from trans-Jovian planets}
\institute{
Hamburger Sternwarte, Universit\"at Hamburg, Gojenbergsweg 112, 
21029 Hamburg, Germany}
\offprints{J.-U.\ Ness}
\mail{jness@hs.uni-hamburg.de}
\date{received \today; accepted}
\maketitle

%
\begin{abstract}

We present an analysis of X-ray observations
of the trans-Jovian planets Saturn, Uranus and Neptune with the ROSAT PSPC in comparison with X-ray
observations of Jupiter. For the first time a marginal X-ray
detection of Saturn was found and 95\% confidence upper limits
for Uranus and Neptune
were obtained. These upper limits show that Jupiter-like X-ray
luminosities can be excluded for all three planets, while they are consistent
assuming intrinsic Saturn-like X-ray luminosities. Similar X-ray production
mechanisms on all trans-Jovian planets can therefore not be ruled out, and
spectral shape and total luminosity observed from Saturn are consistent with
thick-target bremsstrahlung caused by electron precipitation as
occurring in auroral emission from the Earth.\\

\end{abstract}
%

\section{Introduction}
X-ray emission from solar system objects has so far been detected from the 
Earth (Rugge et al. \cite{rug79}, \ Fink et al. \cite{fink88}), 
from the Moon (Schmitt et al. \cite{schmitt91}), 
from several comets (e.g., Lisse et al. \cite{lisse96}, 
Mumma et al. \cite{mumma97}) and from 
Jupiter (e.g., \ Metzger et al. \cite{metzger83}). 
The observed X-ray emission seems to have different physical 
origins in the different objects. The principal X-ray production 
mechanism for Moon and Earth is reflection of solar X-rays;
auroral X-ray emission has been found from the Earth and from Jupiter, and 
similar emission from the outer planets is anticipated.

Aurorae on Earth and Jupiter are generated by charged particles precipitating 
into the atmosphere along the magnetic field lines.
While at Earth the precipitating flux consists of solar wind electrons,
there is strong evidence from the {\it Einstein} observations 
(e.g., \ Metzger et al. \cite{metzger83}) that
the Jovian X-rays are caused by heavy ion precipitation.
Assuming energetic electron precipitation an input power of $\sim10^{15}$ to $\sim10^{16}$~W 
was estimated which seemed to be unreasonably large compared with both
the auroral input power calculated on the basis of the {\it Voyager} observations of the 
UV aurora ($\sim10^{13}-10^{14}$~W) and
with the power estimated to be available in the magnetosphere through mass 
loading in the torus or pitch-angle scattering induced by wave-particle interactions.
From this and from a direct observation of heavy ions in Jupiter's
magnetosphere with the
{\it Voyager} spacecraft, Metzger et al. (\cite{metzger83}) concluded that heavy ion 
precipitation is a reasonable X-ray production process (for references, see Metzger et al. 
\cite{metzger83}). With the {\it Einstein Observatory} Imaging 
Proportional Counter (IPC) pulse height spectrum, both a continuous spectrum resulting 
from bremsstrahlung and a characteristic line emission spectrum from heavy ions, especially 
from oxygen and sulfur, are consistent. 
Because of this inability of the IPC to distinguish between continuous 
emission and line emission, the possibility that the Jovian X-ray emission is due
to bremsstrahlung could not be ruled out, but a
comparison of ROSAT observations in the soft X-ray spectrum with 
model-generated bremsstrahlung and line emission spectra strengthened 
the case for heavy ion precipitation (Waite et al. \cite{waite94}).

X-ray emission from the other outer planets and especially from Saturn is expected 
because of the discovery of magnetospheres by the {\it Voyager} spacecraft 
(e.g. Opp\ \cite{opp80}; Sandel et al.\ \cite{sandel82}) and the observation 
of auroral ultraviolet emission from Saturn at high latitude regions 
(Broadfoot et al.\ \cite{broad81}), from Uranus (Herbert \& Sandel \cite{herb89}) near the poles
and from Neptune (Broadfoot et al. \cite{broad89}).
Broadfoot et al.\ (\cite{broad81}) conclude from their UV observations 
that magnetotail activity on Saturn is more Earth-like and quite different 
from the dominant Io plasma torus mechanism on Jupiter.
If energetic particles are responsible for the observed UV emission,
associated X-ray emission is also expected.

On 1979 December 17 Saturn was observed with the {\it Einstein Observatory} IPC for 
10,850 seconds, but no 
X-ray emission was detected, leading Gilman et al.\ (\cite{gilman86}) to the conclusion that
bremsstrahlung was the more likely X-ray production mechanism for Saturn. 
From this spectral assumption they calculated from the IPC observation 
a $3\sigma$ upper limit for the Saturnian X-ray flux at Earth of 
$1.7~\times~10^{-13}$~erg~cm$^{-2}$~s$^{-1}$. This value has to be compared 
with an expected energy flux at Earth of $8\times 10^{-16}$~erg~cm$^{-2}$~s$^{-1}$, obtained 
from a model calculation (Gilman et al. \cite{gilman86}) based on 
UV observations (Sandel et al.\ \cite{sandel82}) and the 
assumption of thick-target bremsstrahlung at high latitudes.

With the ROSAT position sensitive proportional counter
(PSPC) a more sensitive soft X-Ray observation of Saturn as well as the first 
X-ray observations of Uranus and Neptune have been carried out. The purpose of this paper 
is to present and to analyze these data.

\section{Observation and Analysis}
The outer planets Saturn, Uranus and Neptune were observed with the ROSAT PSPC in the pointing 
mode. Details of the observations such as date, elapsed time, ROSAT 
sequence numbers, number of observation intervals (OBI),
apparent angular size, distance from Earth and other relevant 
items are summarized in Tab. \ref{obs}; for purposes of comparison we also 
analyzed and list a ROSAT PSPC observation of Jupiter, discussed in detail 
by Waite et al. (\cite{waite94}).
\begin{table*}
  \caption[]{Observational details for the outer planets}
  \begin{tabular}{|l|c|c|c||c|}  \hline
&{\bf Saturn}&{\bf Uranus}&{\bf Neptune}&{\bf Jupiter}\\
\hline
Obs\_req\_ID&900160&900130&900132&900013\\
Start\_date&30Apr1992&04Apr1991&03Apr1991&23Apr1991\\
elapsed time[sec]&288960&867377&113915&189971\\
Obs\_time[sec]&5349&10571&17563&5042\\
OBIs&3&7&8&8\\
Ang. diameter&17\arcsec&3\arcsec&2\arcsec&38\arcsec\\
distance from
Earth[AU]&9.9&19.4&30.2&5.2\\
apparent motion&8.8\arcmin&5.12\arcmin&39.7\arcsec&10.5\arcmin\\
\hline
{\it results}&&&&\\
\hline
{\bf soft band}
&&&&\\
source counts+background&22&20&45&211\\
cts expected
from backgr&7.6&21.6&41.4&8.32\\
probability for
no source&$1.7\times 10^{-5}$&0.66&0.31&0\\
energy flux
[erg cm$^{-2}$ s$^{-1}$]&$1.9\times 10^{-14}$&--&--&$2.9\times 10^{-13}$\\
95\% confidence upper limit&&&&\\
\ [cts]&13&30&53&14\\
\ [erg cm$^{-2}$ s$^{-1}$]&--&$5.7\times 10^{-15}$&$4.7\times 10^{-15}$&--\\
extrapolation from Jupiter&&&&\\
\ [erg cm$^{-2}$ s$^{-1}$]&$8.0\times 10^{-14}$&$2.1\times 10^{-14}$&$8.6\times 10^{-15}$&--\\
extrapolation from Saturn&&&&\\
\ [erg cm$^{-2}$ s$^{-1}$]&--&$5.0\times 10^{-15}$&$2.1\times 10^{-15}$&--\\
\hline
{\bf hard band}
&&&&\\
source counts+background&4&25&18&49\\
cts expected
from backgr&2.4&19.4&21.6&1.6\\
probability for
no source&0.22&0.13&0.81&0\\
energy flux
[erg cm$^{-2}$ s$^{-1}$]&--&--&--&$1.9\times 10^{-13}$\\
95\% confidence upper limit&&&&\\
\ [cts]&6&28&30&5\\
\ [erg cm$^{-2}$ s$^{-1}$]&$1.3\times 10^{-14}$&$1.6\times 10^{-14}$&$1.4\times 10^{-14}$&--\\
  \hline
  \end{tabular}
  \label{obs}
\end{table*}
As can be seen from Tab. \ref{obs}, the largest of the planets targeted in these
observations, Saturn, had an apparent size of 17\arcsec\ at the time of the observation. 
Since this is still small compared to the ROSAT PSPC point spread function
(50\% encircled energy is contained within radius of 22\arcsec\ for angles up to
10\arcmin\ with respect to the optical axis),
we will 
treat the data as emission from point sources.
As can be further seen from Tab. \ref{obs}, the elapsed time of the PSPC observations 
was quite long leading to significant motions of the planets during that period. The 
ROSAT standard data processing provides the position of each recorded photon 
with respect to a fixed reference frame.
Since the PSPC is photon counting, we know the arrival time for each 
recorded photon. The planet ephemeris are also known as a function of time, and 
therefore, we can calculate the position shifts $\Delta\alpha$ and $\Delta\delta$ 
to all photons required to correct for the planetary motion. This procedure 
combines all planetary photons into a point source, while photons from 
sources with fixed celestial positions will yield multiple sources 
reflecting the planetary motions.
The thus transformed images were analyzed in the soft energy range taking counts
 in the amplitude channel range from channel 10 to 60 ($\sim 0.1-0.55$~keV) and in
 the hard energy range with channels from 61 to 160($\sim 0.55-1.6$~keV). From the
 count rate we calculated the energy flux using a
conversion factor of $6\times 10^{-12}$~erg~cm$^{-2}$~cts$^{-1}$ for the soft band and $2\times 10^{-11}$~erg~cm$^{-2}$~cts$^{-1}$ for the hard band.
The analysis was carried out in two different ways.
The first method simply consists of placing a square box
on the planet's position and comparing the source 
box counts with the background counts determined from a much larger box 
placed in the vicinity of the source box but containing no sources. 
For practical purposes we treat Saturn as a point source, since resolving
17\arcsec\ requires an extremely high signal to noise ratio.
In order to pick up as many source counts as possible while keeping the
background low, we choose a box size of $1.5\arcmin\times 1.5\arcmin$.
For the soft X-ray range this source box contains 83.6\% of all source photons.
This was empirically determined from the supersoft white dwarf HZ43; the energy fluxes 
given in Tab. \ref{obs} for the soft energy range are corrected by this value.
In the soft energy range we thus have an expected source box count of $7.6~\pm~0.1$ at the 
position of Saturn from background alone, but we find 22 counts,
concentrated towards the center of the source box as expected for a point-like
source.
The probability of measuring 22 counts or more with only 7.6 counts are expected is 
$1.7\times 10^{-5}$, assuming Poisson statistics.
The corresponding numbers in the hard energy range are $2.4~\pm~0.1$ expected counts 
with 4 counts actually recorded. The probability for measuring at least four counts 
assuming no source is 22\%. Clearly, the signal recorded in the hard band is consistent 
with a background fluctuation, while in the soft band a significant excess is seen.
These numbers are recorded in Tab. 1 for all target planets (as well as Jupiter) for both the
soft and hard energy band.

Our second approach consists of applying the maximum likelihood detection technique 
described by Cruddace et al.\ (\cite{crudd88}) to the transformed images. This procedure
results in a source existence maximum likelihood of 3.1 at Saturn's position for the soft 
energy range and again no detection in the hard energy range.
Clearly, the source existence likelihood is low. In judging the significance level 
one must however keep in mind that X-ray emission was searched for at only one position. 
A confirmation of this detection by another satellite measurement thus is highly desirable.
Accepting for the time being the ROSAT detection of Saturn as real,
we find a count rate of
2.7$\times 10^{-3}$~cts/s which corresponds to an incident 
energy flux of $1.9\times 10^{-14}$~erg~cm$^{-2}$~s$^{-1}$.

It is obvious from Tab.~\ref{obs} that no detections of Uranus nor Neptune have been obtained.
From the computed 95\% confidence count upper limits we calculated flux upper 
limits in the soft and hard energy bands, which are also listed in 
Tab.~\ref{obs}.

\section{Discussion}
\label{discussion}
Our analysis of the ROSAT PSPC data on trans-Jovian planets
yields a marginal X-ray detection of Saturn, while only 
upper limits could be obtained for Uranus and Neptune. These upper limits are sensitive
in the sense that X-ray emission at the level of Jupiter from Uranus and Neptune
 would have been detected. However, the upper limits are consistent assuming
intrinsic Saturn-like X-ray luminosities for Uranus and Neptune.
Therefore, similar X-ray production mechanisms on all trans-Jovian planets can certainly
not be ruled out from the currently available observations. On the other hand, it also
appears that Jupiter is rather unique with regard to its X-ray luminosity 
(Gilman et al.\ \cite{gilman86}, Waite et al.\ \cite{waite94}). 
A possible explanation for the X-ray production on the trans-Jovian 
planets is thick-target bremsstrahlung caused by electron precipitation, i.e.,
the process which is also responsible for the generation of aurorae on the Earth.
Gilman et al. (\cite{gilman86}) (cf. Tab. \ref{comp}) expect an X-ray flux from Saturn
of $8\times 10^{-16}$ erg~cm$^{-2}$~s$^{-1}$ from thick-target bremsstrahlung, based on 
the observed UV-flux and the assumption of a power-law electron distribution function.
An even lower value is however obtained from the measured electron flux in Saturn's
outer magnetosphere with the {\it Voyager} spacecraft, assuming, at high
energies, an exponential in electron speed (Barbosa \cite{bosa90}).
Saturn's energy flux obtained from our PSPC observation exceeds these expected
fluxes by more than one order of magnitude.
This might be either due to an elevated electron flux at the time of the observation
or to other X-ray production mechanisms in addition to bremsstrahlung.
Since the Saturnian system does not contain a volcanically active moon like Io,
for example, it is unclear how heavy ions can be efficiently inserted into
Saturn's magnetosphere.

\begin{table}
 \caption[]{Comparison of {\it Einstein} and ROSAT observations of Saturn}
 \begin{tabular}{l|c}  \hline
&Flux [erg~cm$^{-2}$~s$^{-1}$]\\
\hline
{\it Einstein} observation&\\
$3\sigma$ upper limit for&\\
\ Bremsstrahlung mechanism&$1.7\times 10^{-13}$\\
\ same process as Jupiter&$5\times 10^{-13}$\\
\hline
model calculation by&\\
\ Gilman et al. (\cite{gilman86})&$8\times 10^{-16}$\\
\hline
our result from ROSAT&$1.9\times 10^{-14}$\\
\hline
  \end{tabular}
  \label{comp}
\end{table}
Support for the assumption of thick-target bremsstrahlung comes from the observed spectral
signatures of the X-ray detection of Saturn, which was detected only in the soft energy band (cf. Tab. 1)
 as well as from a comparison of aurorae on Earth observed
with the very same instrument, i.e., the ROSAT PSPC, with which Saturn was observed and analyzed also in the soft energy band.
Freyberg (\cite{frey94}) discusses a number of PSPC observations which show a strong enhancement
in the diffuse background count rate and which can be traced back to auroral activity and/or
geomagnetic storms. The relevant data are summarized in Tab. 3. A specific example is
the ROSAT observation CA150057, which showed a significantly enhanced background (almost
exclusively in the soft energy band) in one observation interval when ROSAT traversed the region
south of Greenland. This elevation in count rate is interpreted as
auroral X-ray emission due to bremsstrahlung at the Earth's atmosphere near the northern radiation belt.
An even more extreme case is the ROSAT PSPC observation WG700232 during which 
an intense geomagnetic storm took place. In the 27$^{\rm th}$ observation interval of this specific data set
the PSPC count rate rose up to a value of more than 2300 counts per second when the PSPC was turned 
off and went in safe mode; also in this case the count rate was highly time variable and consisted
of very soft photons. In both cases we can determine the observed PSPC intensity (in units
of counts/sec/arcsec$^2$) by subtracting the observed background from observation intervals 
unaffected by auroral emission; the results are listed in Tab. 3.

\begin{table}
  \caption[]{Comparison of auroral X-ray emission from Earth (Freyberg (\cite{frey94})) and from Saturn (soft band).}
  \begin{tabular}{|l|c|c||c|}  \hline
&CA150057&WG700232&Saturn\\
Countrate+backgr&&&\\
\ [cts~sec$^{-1}$]&75.11&$>$2300&$4.1\cdot 10^{-3}$\\
backgr [cts~sec$^{-1}$]&13.58&28&$1.4\cdot 10^{-3}$\\
\hline
area [arcsec$^{2}$]&$4.1\cdot 10^{7}$&$4.1\cdot 10^{7}$&227\\
intensity&&&\\
\ [cts~sec$^{-1}$~arcsec$^{-2}$]&$1.5\cdot 10^{-6}$&$>5.6\cdot 10^{-5}$&$1.2\cdot 10^{-5}$\\
\hline
  \end{tabular}
  \label{aurora}
\end{table}

A comparison of the observed PSPC intensities of aurorae on Earth in the soft
energy band with that of Saturn shows that
the former reach values that can easily account for the observed emission from Saturn. Note
in particular that the X-ray emission during a geomagnetic storm may be much higher. If therefore
-- as appears likely -- the X-ray emission from Saturn is restricted to its auroral belts, the resulting
X-ray intensities may still be in the range of X-ray intensities observed in geomagnetic storms on Earth. 

In summary, we can state that Jupiter's X-ray emission among the solar system planets appears unique
in terms of total luminosity and possibly also in terms of spectral shape. No other planet has an intrinsic
X-ray luminosity as high as Jupiter, and furthermore, Waite et al. (\cite{waite94}) suggest that
its X-ray emission seems to be dominated by lines
rather than continuum emission. The obtained X-ray detection of Saturn appears to be consistent
both in strength and spectral shape with thick-target bremsstrahlung as occurring
in auroral emission on Earth, but the observed luminosity implies rather high
electron fluxes. It is clearly highly desirable to obtain a high angular resolution X-ray image
of Saturn in order to confirm, first, the X-ray detection obtained with the ROSAT PSPC and, second, to
study the spatial distribution of the X-ray emission on Saturn's surface, which is expected to be
concentrated in Saturn's auroral belts.

\begin{acknowledgements}
J.-U.N. acknowledges financial support from Deutsches Zentrum f\"ur Luft- und
Raumfahrt e.V. (DLR) under 50OR98010.
\end{acknowledgements}



\begin{thebibliography}{}
 
\bibitem[1990]{bosa90}
   Barbosa D.D., 1990, Planet. Space Sci, 38, 1295
\bibitem[1981]{broad81}
   Broadfoot A.L., et al., 1981, Science 212, 206
\bibitem[1989]{broad89}
   Broadfoot A.L., et al., 1989, Science 246, 1459
\bibitem[1988]{crudd88}
   Cruddace R.G., Hasinger G.R., Schmitt J.H.M.M., ''Astronomy from large databases: Scientific objectives and methodological approaches'' in: Proceedings of the Conference, Garching, Federal Republic of Germany, Oct. 12-14, 1987 (A89-27176 10-82).
Garching, Federal Republic of Germany, European Southern Observatory, 1988, p. 177.
\bibitem[1988]{fink88}
   Fink H.H., et al., 1988, A\&A 193, 345
\bibitem[1994]{frey94}
   Freyberg M., 1994, PhD thesis Ludwig-Maximilians-Universit\"at M\"unchen
\bibitem[1986]{gilman86}
   Gilman D.A., et al., 1986, ApJ 300, 453
\bibitem[1989]{herb89}
   Herbert F., Sandel B.R., 1989, EOS 70,1174
\bibitem[1983]{hill83}
   Hill T., Dessler A.J., Goertz C.K., 1983, in {\it Physics of the Jovian Magnetosphere}, ed. A.J. Dessler (Cambridge: Cambridge University Press), p.365
\bibitem[1996]{lisse96}
   Lisse C.M., et al., 1996, Science 274, 205L
\bibitem[1997]{mumma97}
   Mumma M.J., et al., 1997, ApJ Letters 491, L125
\bibitem[1983]{metzger83}
   Metzger A.E., et al., 1983, J. Geophys. Res. 88, 7731
\bibitem[1980]{opp80}
   Opp A.G., 1980, Science 207, 400
\bibitem[1979]{rug79}
   Rugge H.R., McKenzie D.L., Charles P.A., 1979, Space Research XIX, 243
\bibitem[1982]{sandel82}
   Sandel B.R., et al., 1982, Science 215, 548
\bibitem[1991]{schmitt91}
   Schmitt J.H.M.M., et al., 1991, Nature 349, 583 
\bibitem[1981]{thorne81}
   Thorne R.M., 1981, Geophys. Res. Letters 8, 509
\bibitem[1994]{waite94}
   Waite J.H., et al., 1994, J. Geophys. Res. 99, 14799
\end{thebibliography}
\end{document}